\documentclass{raa}           

\usepackage{graphicx,times}
\usepackage{natbib}
\usepackage{amssymb,amsmath}
\usepackage{graphicx}
\usepackage{subcaption}
\bibpunct{(}{)}{;}{a}{}{,}

\usepackage[pagebackref=true]{hyperref}

\begin{document}

   \title{The Abundance Origin Of A Highly $r$-process-Enhanced r-II Star: LAMOST J020623.21+494127.9}

 \volnopage{ {\bf 20XX} Vol.\ {\bf X} No. {\bf XX}, 000--000}
   \setcounter{page}{1}

   \author{Muhammad Zeshan Ashraf 
   \inst{1}, Wenyuan Cui\inst{1}
      ,  Hongjie Li\inst{2}
   }

   \institute{ College of Physics, Hebei Normal University, Shijiazhuang, 050024, 
China; {cuiwenyuan@hebtu.edu.cn}\\
        \and
             College of Sciences, Hebei University of Science and Technology, Shijiazhuang, 050018, China\\
\vs \no
   {\small Received 20XX Month Day; accepted 20XX Month Day}
}

\abstract{ 
       Object LAMOST J020623.21+494127.9 (program star) in the thin disk of the Milky Way (MW) is reported as a highly $r$-process-enhanced (RPE) r-II star with [Eu/Fe]= +1.32 \, \text{and} \, [Fe/H]= -0.54. The chemical profile of the star reflects the intrinsic composition of the gas cloud present at its birth. Using an abundance decomposition method, we fit 25 elements from the abundance dataset, including 10 heavy neutron-capture elements. We explore the astrophysical origin of the elements in this star through its abundance ratios and component ratios. We find that the contributions from the massive stars played a significant role in the production of light elements in the program star. Our analysis reveals that the heavy neutron-capture elements are produced purely by the main $r$-process. However, the adopted main $r$-process model does not adequately fit the observed data, suggesting another main $r$-process pattern may exist. \\
       \keywords{ 
           Nucleosynthesis, neutron-capture elements, main $r$-process, weak $r$-process, stellar abundances, chemical evolution 
       }
   }

   \authorrunning{Zeshan et al. }            
   \titlerunning{Zeshan et al.}  
   \maketitle
%
\section{Introduction}           
\label{sect:intro}

The elements heavier than iron are mainly produced by the slow ($s$) and rapid ($r$) neutron-capture processes. The $s$-process is divided into two subclasses. The weak $s$-process is responsible for producing light neutron-capture elements below $A=90$ and takes place in the core He-burning and shell C-burning phases of massive stars (\citealt{Woosley+etal+2002}). The elements in the mass number range \(88 < A < 208\) are generally produced by the main $s$-process (\citealt{Arlandini+etal+1999}), and its possible sites are asymptotic giant branch (AGB) stars with low to intermediate mass (\citealt{Gallino+etal+1998, Bisterzo+etal+2011}). The $r$-process also comprises two subclasses: the weak $r$- and the main $r$-processes. The weak $r$-process, which is sometimes also called the "lighter element primary process," can only produce light neutron-capture elements with mass number \(A < 130\). Its sites are probably Type II supernovae (SNe II) with progenitors of \(M\geq 10 M_\odot\) (\citealt{Travaglio+etal+2004, Ishimaru+etal+2005}). The main $r$-process can produce both light and heavy neutron-capture elements. Many possible sites have been suggested for the main $r$-process, including core-collapse supernovae (CCSNe) and binary neutron star mergers (NSMs) (\citealt{Burbidge+etal+1957, MacFadyen+Woosley+1999, Winteler+etal+2012, Abbott+etal+2017}). However, there also exist other suggestions for main $r$-process sources, like Magneto-rotational supernovae (MRSNe) and Collapsars (\citealt{Thielemann+etal+2023}.
The exact origin of the main $r$-process is still not confirmed (\citealt{Sneden+etal+2008, Thielemann+etal+2011, Farouqi+etal+2022}).

The $r$-process enhanced (RPE) stars provide important clues about the nature of the $r$-process, which may be essential for understanding the astrophysical origins of the $r$-process. RPE stars are divided into two categories, namely r-I \((0.3 \leq [\text{Eu/Fe}] \leq 0.7, [\text{Ba/Eu}] < 0)\), and r-II \([\text{Eu/Fe}] > 0.7, [\text{Ba/Eu}] < 0\) (\citealt{Holmbeck+etal+2020, Frebel+Ji+2022}). Europium (Eu) is considered a primary indicator of $r$-process enrichment because the $r$-process primarily determines its abundance and can be easily measured from high-resolution optical spectra. The limits on barium (Ba) are considered to eliminate the contribution from the $s$-process. It has been reported that the abundance pattern of heavy elements \(Z \geq\ 56\) in both r-I and r-II stars aligns well with the solar-scaled $r$-process pattern, indicating that the main $r$-process is a universal mechanism of nucleosynthesis (\citealt{Burris+etal+2000, Ivans+etal+2006, Frebel+etal+2007, Roederer+etal+2023}).
The r-II stars are very important for investigating the pure $r$-process signature because they are less affected by the $s$-process (\citealt{Frebel+2010}). These stars are generally metal-poor \([\text{Fe/H}] < -2.5\), and they are found in the Milky Way halo and dwarf galaxies (\citealt{Sneden+etal+1996, Cayrel+etal+2001, Barklem+etal+2005, Ji+etal+2016}), while r-II stars in the metallicity range (\(-2 < [\text{Fe/H}] < -1\)) are rare. Many r-II stars have been discovered over the past two decades, but only three r-II stars with \(-2 < [\text{Fe/H}] < -1\) have been identified in the Milky Way: 2MASS 18174532-3353235 \([\text{Fe/H}] = -1.67, [\text{Eu/Fe}] = 0.99\), J1802-4404 \([\text{Fe/H}] = -1.55, [\text{Eu/Fe}] = 1.05\), and HD 222925 \([\text{Fe/H}] = -1.46, [\text{Eu/Fe}] = 1.32\) (\citealt{Johnson+etal+2013, Hansen+etal+2018, Roederer+etal+2018, Roederer+etal+2022}). 2MASS 18174532-3353235 is a bulge star, while J1802-4404 and HD 222925 belong to the halo of the Milky Way. Many moderately r-II stars have also been discovered in dwarf galaxies, with the ultra-faint dwarf galaxy Reticulum II, the Fornax dwarf galaxy, and the Ursa Minor dwarf galaxy being particularly significant. Two stars, COS 82 and J033457-540531, are highly $r$-process enriched and were found in the Ursa Minor and Reticulum II dwarf galaxies, respectively (\citealt{Aoki+etal+2007, Ji+etal+2016}). COS 82 has \([\text{Fe/H}] = -1.42\) and \([\text{Eu/Fe}] = 1.49\), while J033457-540531 has \([\text{Fe/H}] \approx -2\) and a highly enriched $r$-process abundance \([\text{Eu/Fe}] = 1.7\). The study of RPE and moderately metal-poor r-II stars offers valuable insights into understanding the nature of the main $r$-process.

LAMOST J020623.21+494127.9 (program star) is a highly RPE (r-II) star with \([\text{Eu/Fe}] = +1.32\) and \([\text{Ba/Eu}] = -0.95\), discovered in the thin disk of the MW (\citealt{Xie+etal+2024}). This star exhibits unusually high metallicity \([\text{Fe/H}] = -0.54\) compared to most other highly RPE stars, and it holds the highest measured abundance ratio of Eu to H \([\text{Eu/H}] = +0.78\). It deserves special attention because it is relatively more metal-rich than previously observed r-II stars. Based on its kinematics, this star is classified as a thin-disk star and does not appear to belong to any known stellar stream or dwarf galaxy (\citealt{Xie+etal+2024}).

As suggested by \citet{Xie+etal+2024}, the program star does not exhibit significant radial-velocity variations, indicating that it is unlikely to be part of a binary system. This implies that the star has not undergone $s$-process enrichment via mass transfer from a binary companion in the AGB phase. The observed \([\text{Ba/Eu}] = -0.95\) ratio further supports this, as it indicates minimal $s$-process contribution to the enrichment of neutron-capture elements. This work will elaborate on the origin of heavy neutron-capture elements in the program star, using its observed abundances and a multiple-component model.

Even within the relatively uniform structure of the Milky Way’s thin disk, the distribution of $r$-process material is not uniform. Therefore, it is suggested that this program star was likely born in a region enriched with $r$-process elements from either a neutron star merger or a core-collapse supernova \citep{Xie+etal+2024}.  However, further investigation is needed to clarify this hypothesis. These considerations motivated us to undertake a comprehensive analysis to explore the astrophysical origin of the elements in this program star, with a particular focus on neutron-capture elements.

In this work, Section \ref{sec:2} discusses the acquisition of the observed element abundance data of this program star and focuses on exploring the astrophysical origins of the elements, by fitting the observed abundances using a multiple-component model. Section \ref{sec:3} analyzes the formation and evolutionary history of this program star. The conclusion is presented in Section \ref{sec:4}.

\section{The Formation and Evolution of Elements in the Program star}
\label{sec:2}
The abundances used in this work were retrieved from the work by \citet{Xie+etal+2024}. They selected this program star for observation using the Large Sky Area Multi-Object Fiber Spectroscopic Telescope through a medium-resolution (R$\sim$7500) spectroscopic survey (\citealt{Cui+etal+2012, Zhao+etal+2012, Liu+etal+2020, Yan+etal+2022}). \citet{Xie+etal+2024} observed its high-resolution spectrum with the High Optical Resolution Spectrograph (HORuS) installed on the Gran Telescopio Canarias (GTC). They employed a spectrum-synthesis approach for the abundance determination, utilizing the line list and atomic data from \citet{Zhao+etal+2016} and \citet{Roederer+etal+2018}.
\subsection{Parametric model and calculations}
The elements in stars are not typically the result of a single nucleosynthetic event but are instead produced through multiple mechanisms. Elements with atomic numbers \(Z < 30\) are generally synthesized in supernova explosions, while both light and heavy neutron-capture elements are formed through the $s$- and/or $r$-processes. We utilize an abundance decomposition approach to explore the astrophysical origins of the elements in this program star. From the abundance dataset given by \citet{Xie+etal+2024}, we fit 25 elements, of which 10 are heavy neutron-capture elements. We adopt the multiple-component model from \citet{Han+etal+2020} to fit the observed abundances of this program star, providing deeper insights into the processes responsible for its enrichment. The multiple components model is represented as:
\begin{equation}
N_i =(C_{r,m} N_{i,r,m}+ C_{pri} N_{i,pri}+C_{s,m} N_{i,s,m}+C_{sec}N_{i,sec} +C_{Ia} N_{i,Ia}) \times 10^{[\text{Fe/H}]}, \
\label{eq:1}
\end{equation}
Where \(N_i\) indicates the abundance of the \(i\)-th element. The terms \(N_{i,r,m}\), \(N_{i,pri}\), \(N_{i,s,m}\), \(N_{i, sec}\), and \(N_{i, Ia}\) represent the abundances of the main $r$-process, primary process, main $s$-process, secondary process, and Type Ia supernovae (SNe Ia), respectively. The coefficients \(C_{r,m}\), \(C_{pri}\), \(C_{s,m}\), \(C_{sec}\), and \(C_{Ia}\) correspond to the respective component contributions.

Massive stars (with masses \(M\geq 10 M_\odot\)) during hydrostatic burning produce primary light and iron-peak elements, while Type II supernovae with progenitor masses \(M\geq 10 M_\odot\) generate weak $r$-process elements. Since massive stars eject primary light, iron-peak, and weak $r$-process elements, these can be combined into one called “primary component.” Hydrostatic burning in massive stars also produces secondary light and iron-peak elements, while core helium burning and shell carbon burning in these stars yield weak $s$-process elements. Therefore, the secondary light, iron-peak, and weak $s$-process elements can be grouped into the “secondary component” (\citealt{Li+etal+2013b}). It is important to note that the abundances of secondary elements are only observable at higher metallicities, as their yields decrease with decreasing metallicity. The abundances of the secondary component \(N_{i,sec}\) are adopted from \citet{Li+etal+2013b}, whereas the abundances \(N_{i,r,m}\) and \(N_{i,pri}\) are taken from \citet{Li+etal+2013a}. The abundances \(N_{i, Ia}\) are adopted from \citet{Timmes+etal+1995}, with updated values for Fe, Cu, and Zn from \citet{Mishenina+etal+2002}. It has been reported that SNe Ia events cannot occur for progenitors with low metallicity \([ \text{Fe/H} ] < -1.0\) (\citealt{Kobayashi+etal+1998}). This suggests that SNe Ia events are restricted to higher-metallicity environments.

From observations, \citet{Xie+etal+2024} reported that this program star exhibits a strong enhancement of Eu (\([ \text{Eu/Fe} ] = +1.32\)) and a moderate enhancement of Ba (\([ \text{Ba/Fe} ] = +0.37\)). The low \([ \text{Ba/Eu} ]\) ratio of \(-0.95\) indicates that the heavy neutron-capture elements in this star are not the result of the main $s$-process, but rather the main $r$-process. To test this argument, the main $s$-process abundances \( N_{i,s,m} \) in Equation (\ref{eq:1}) are taken from the abundances produced by 1.5 \( M_{\odot} \) AGB stars with \([ \text{Fe/H} ] = -0.6\) which is given by \citet{Busso+etal+2001}. We can derive the five component coefficients in Equation (\ref{eq:1}) by comparing the calculated abundance \(N_{i, cal}\) and observed abundance \(N_{i, obs}\) and seeking the smallest $\chi^2$ which is defined as:
\begin{equation}
\chi^2=\sum_{i=1}^{n}\frac{\big(logN_{i,obs}-logN_{i,cal}\big)^2}{\big(\Delta logN_{i,obs}\big)^2 \big(K-K_{free}\big)},
\label{eq:2}
\end{equation}
Where \( \Delta \log N_{i, \text{obs}} \) represents the observed abundance errors of \( i \)-th element, which are adopted from \citet{Xie+etal+2024}. \( K \) and \( K_{\text{free}} \) denote the total number of elements studied and the number of free parameters, respectively. In this case, the values of \( K \) and \( K_{\text{free}} \) are 25 and 5, respectively. 

In the calculation, we obtain a minimum value of \( \chi^2 = 1.31 \), and the component coefficients are \( C_{r,m} = 18.05 \), \( C_{\text{pri}} = 1.36 \), \( C_{s,m} = 0 \), \( C_{\text{sec}} = 1.18 \), and \( C_{\text{Ia}} = 0.56 \). It is evident from the calculation that the value of \( C_{r,m} \) is significantly higher than the rest of the component coefficients, indicating that the heavy neutron-capture elements in this program star are predominantly produced by the main $r$-process. The calculated value of \( C_{s,m} \) is zero, supporting the claim of \citet{Xie+etal+2024} that the $s$-process is not responsible for the production of heavy neutron-capture elements in this star. This makes our program star a precious sample to study the pure $r$-process signatures. The value of \(\chi^2\) in the multiple-component model is sensitive to both the observed abundance errors and the errors inherent in the model itself. According to Equation (\ref{eq:2}), when the value of \( K \) is large, the \(\chi^2\) quantifies the comparison of the relative offsets (\(\Delta \log N = \log N_{i,\text{obs}} - \log N_{i,\text{cal}}\)) and the observed errors \( \Delta \log N_{i, \text{obs}} \). If the \(\chi^2\) value is close to 1, or of the order of unity, the relative offsets are nearly equal to the observed errors, indicating good agreement between the observed and model-predicted abundances. In our case, \(\chi^2\) value is 1.31, which can be considered a satisfactory result.

\begin{figure}[http]
    \centering
    \includegraphics[width=0.8\linewidth]{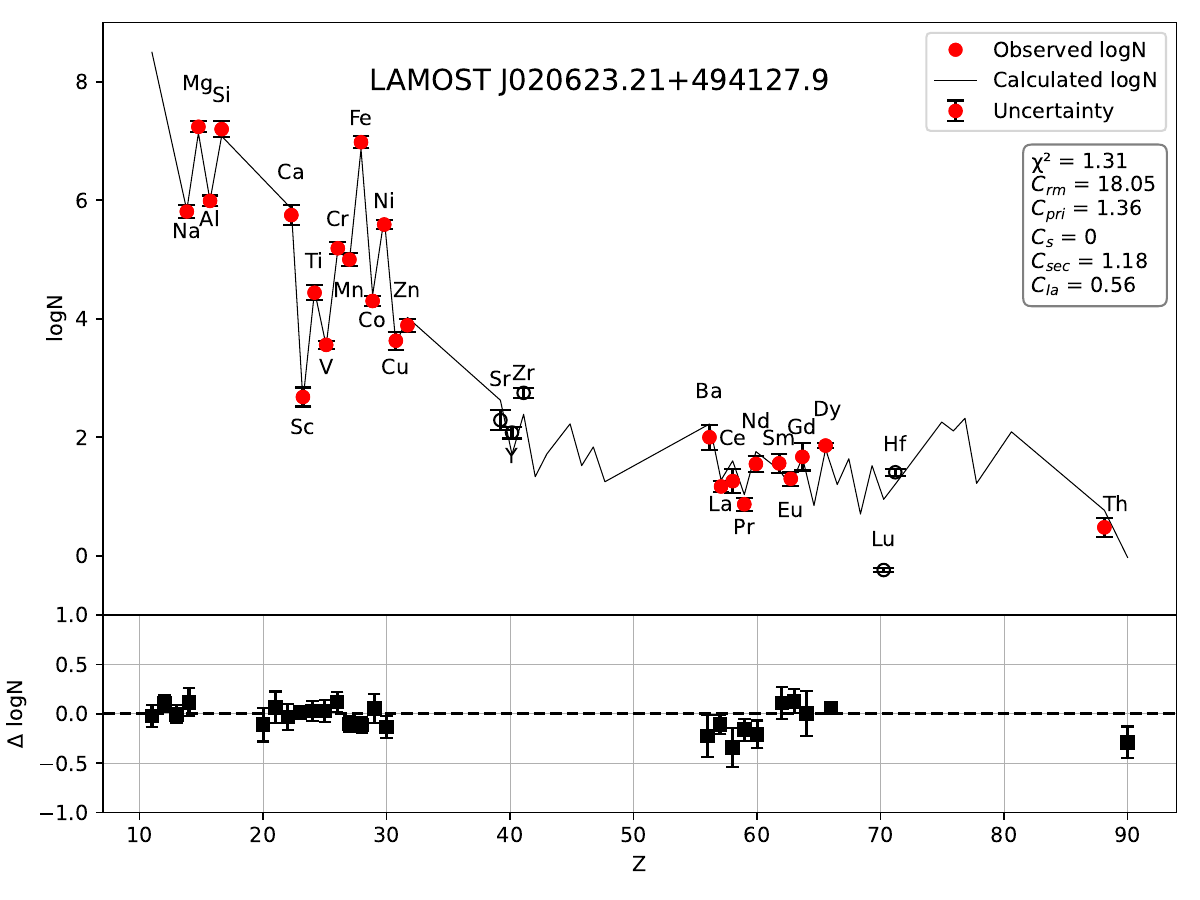}
    \caption{Top panels: The best fit of the calculated abundances in the program star (the solid line) and observed abundances (filled red circles). The hollow circles represent elements that are not included in the fitting. Bottom panels: the relative offsets (\(\Delta \log N = \log N_{i,\text{obs}} - \log N_{i,\text{cal}}\)) and the observed uncertainties.}
    \label{fig:enter-labe1}
\end{figure}

The best fit between the observed and calculated abundances of this program star is shown in Figure \ref{fig:enter-labe1}. The observed abundances are represented by red-filled circles, while the calculated abundances are represented by the solid black line in the top panel. The relative offsets (\(\Delta \log N = \log N_{i,\text{obs}} - \log N_{i,\text{cal}}\)), along with the observed uncertainties, are illustrated in the bottom panel. The abundances of Sr, Y, Zr, Lu, and Hf in the program star are excluded from the fitting process because the adopted component model fails to reproduce them accurately. The observed abundances of these elements are depicted by hollow circles, accompanied by their corresponding observational uncertainties. It is evident that the abundances of all elements, except these, are well explained. Therefore, the abundances of the remaining elements are sufficient to constrain the five component coefficients. According to \citet{Xie+etal+2024}, the abundance pattern of heavy elements in this star is not perfectly consistent with the solar $r$-process. Specifically, the abundances of elements Sr, Y, Ce, Pr, and Nd are lower than expected compared to the solar $r$-process pattern. This distinction highlights the likelihood of diverse origins for the $r$-process. 
The investigation of component ratios of each process can help us to explain the abundance features of Sr, Y, and Zr in this program star. For that purpose, the component coefficients derived from Equation (\ref{eq:1}) are used to calculate the theoretical abundances of Sr, Y, and Zr.
\begin{equation}
\left[ \frac{E_i}{E_j} \right]_k = \log \left( \frac{N_{i,k}}{N_{j,k}} \right) - \log \left( \frac{N_{i,\odot}}{N_{j,\odot}} \right),
\label{eq:3}
\end{equation}
In Equation (\ref{eq:3}), \( E_i \) and \( E_j \) denote the elements being compared, such as Sr, Y, or Zr. The subscript \( k \) indicates the specific process under consideration, such as the main $r$-process, primary process, or secondary process. The terms \( N_{i,k} \) and \( N_{j,k} \) represent the number abundances of elements \( E_i \) and \( E_j \) in process \( k \), while \( N_{i,\odot} \) and \( N_{j,\odot} \) represent the solar abundances of these elements. The calculated results are shown in table (\ref{tab:1}).
\begin{table}[h!]
\centering
\caption{Comparison of abundance ratios [Sr/Y], [Y/Zr], and [Sr/Zr] for different processes}
 \renewcommand{\arraystretch}{1.5}
\begin{tabular}{|c|c|c|c|}
        \hline
        \textbf{Process} & \textbf{[Sr/Y]$_k$} & \textbf{[Y/Zr]$_k$} & \textbf{[Sr/Zr]$_k$} \\
        \hline
        Main $r$-process & 0.23 & -0.30 & -0.06 \\
        \hline
        Primary process & 0.07 & -0.29 & -0.21 \\
        \hline
        Secondary process & 0.10 & 0.56 & 0.66 \\
        \hline
        Observed & -0.50 & -0.26 & -0.76 \\
        \hline
    \end{tabular}
\label{tab:1}
\end{table}

Since the observed [Sr/Y]$_{\text{obs}} = -0.50$ is less than the calculated values for each process, it is evident that Sr is generated in far smaller amounts than Y. The unique environmental conditions of this program star may account for this discrepancy. Conversely, the observed [Y/Zr]$_{\text{obs}} = -0.26$ closely matches with the calculated values from the primary and main $r$-processes, which are -0.29 and -0.30, respectively. This resemblance indicates that the production of Y and Zr in this program star could be attributed to either the primary process, the main $r$-process, or a combination of both. However, it is challenging to distinguish the dominant source based purely on this ratio, as the predictions of these processes closely resemble each other.  A more detailed analysis of additional abundance patterns or enrichment signatures may be required to better understand the relative contributions of different $r$-process pathways.
 Furthermore, the observed [Sr/Zr]$_{\text{obs}}$ is lower than the calculated values for the three processes. These findings indicate that there are conditions in this star that enhance the formation of Zr relative to Sr, suggesting that adopted nucleosynthesis models may not fully account for the environmental factors affecting this specific star's elemental abundances. We will further discuss the possible origins of the lighter and heavier neutron-capture elements in the next section.
\subsection{The astrophysical origins of elements in the Program star}
\begin{figure}[http]
    \centering
    \includegraphics[width=0.8\linewidth]{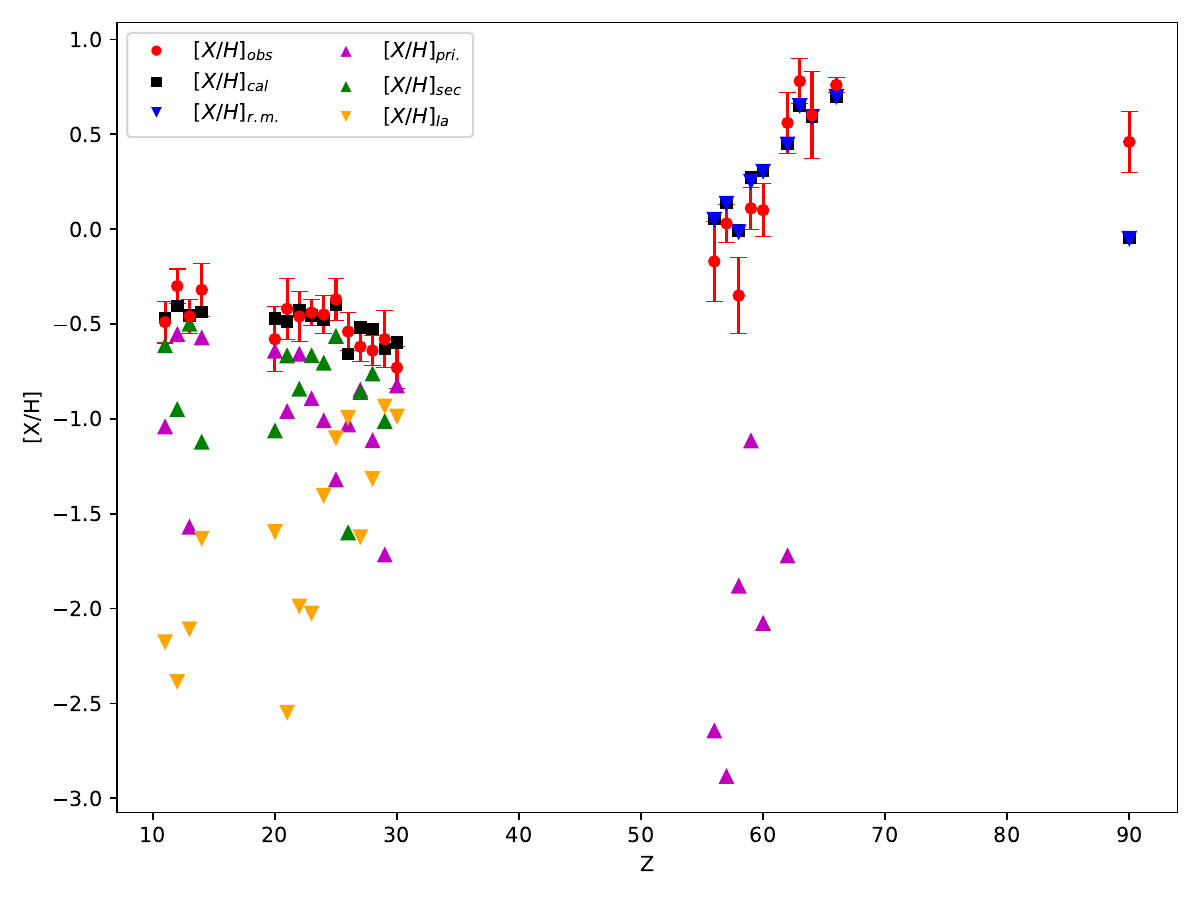}
\caption{The observed and calculated abundance ratios, and component ratios of our program star. The component ratios include the ratios of the main $r$-process, primary process, secondary process, and SNe Ia components, respectively.}
    \label{fig:enter-labe2}
\end{figure}

\begin{figure}[htbp]
    \centering
    \begin{subfigure}{0.33\textwidth}
        \centering
        \includegraphics[width=\textwidth]{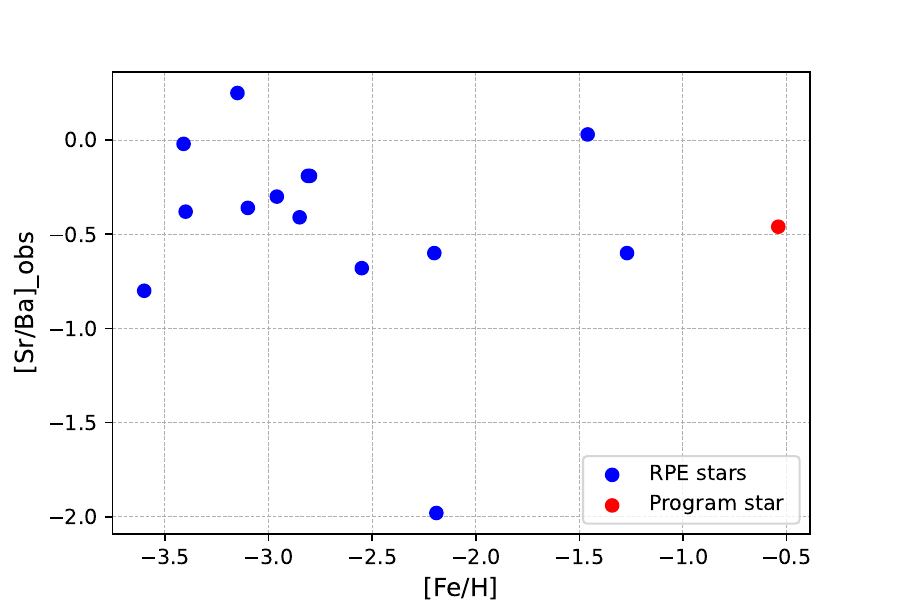}  
        \caption{[Sr/Ba] vs [Fe/H]}
        \label{fig:subfig3a}
    \end{subfigure}
    \begin{subfigure}{0.33\textwidth}
        \centering
        \includegraphics[width=\textwidth]{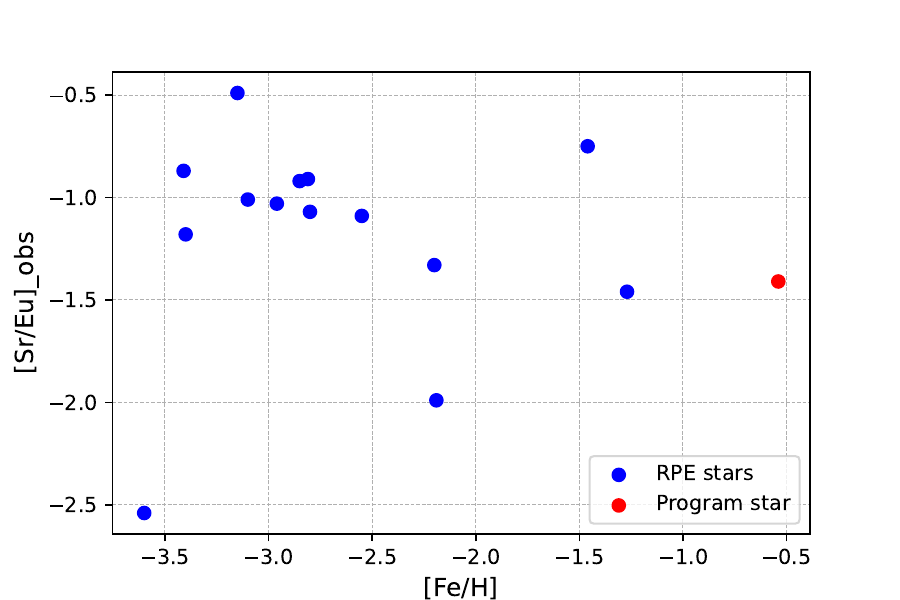}  
        \caption{[Sr/Eu] vs [Fe/H]}
        \label{fig:subfig3b}
    \end{subfigure}
    \begin{subfigure}{0.33\textwidth}
        \centering
        \includegraphics[width=\textwidth]{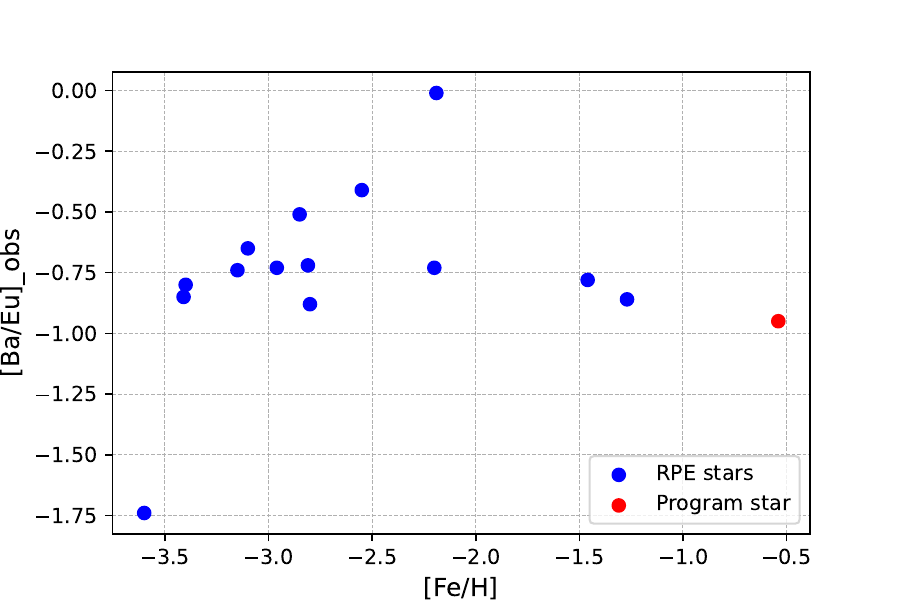}  
        \caption{[Ba/Eu] vs [Fe/H]}
        \label{fig:subfig3c}
    \end{subfigure}
    
    \begin{subfigure}{0.33\textwidth}
        \centering
        \includegraphics[width=\textwidth]{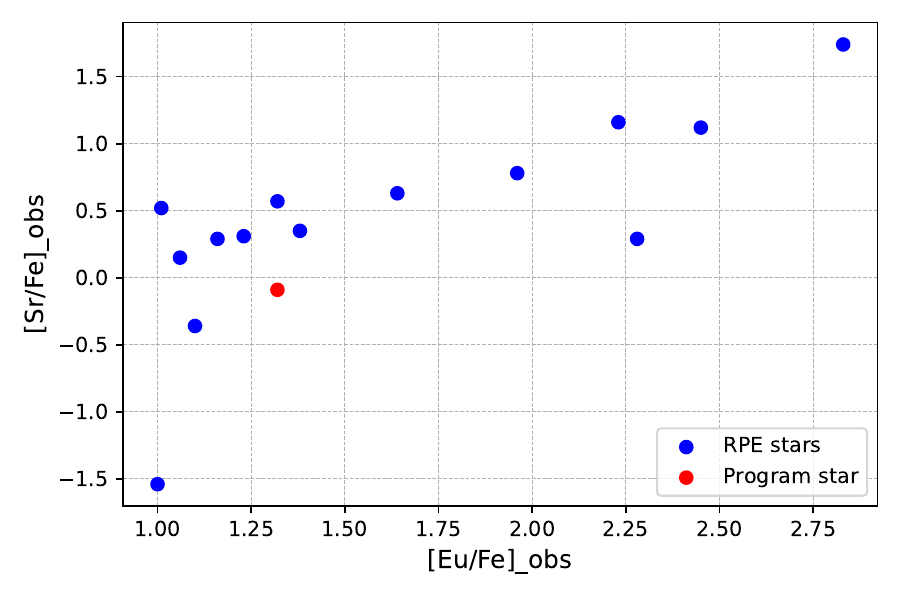}  
        \caption{[Sr/Fe] vs [Eu/Fe]}
        \label{fig:subfig3d}
    \end{subfigure}
    \begin{subfigure}{0.33\textwidth}
        \centering
        \includegraphics[width=\textwidth]{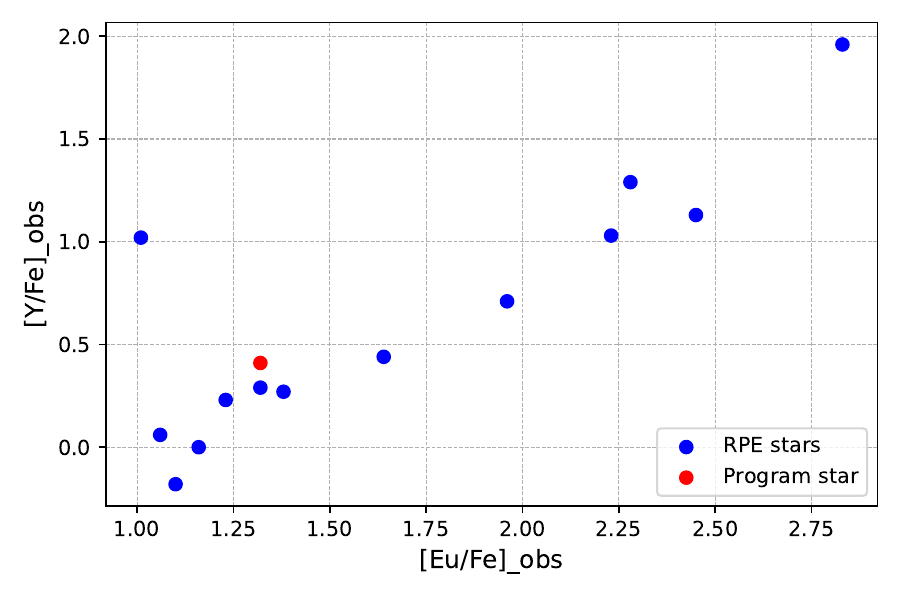}  
        \caption{[Y/Fe] vs [Eu/Fe]}
        \label{fig:subfig3e}
    \end{subfigure}
    \begin{subfigure}{0.33\textwidth}
        \centering
        \includegraphics[width=\textwidth]{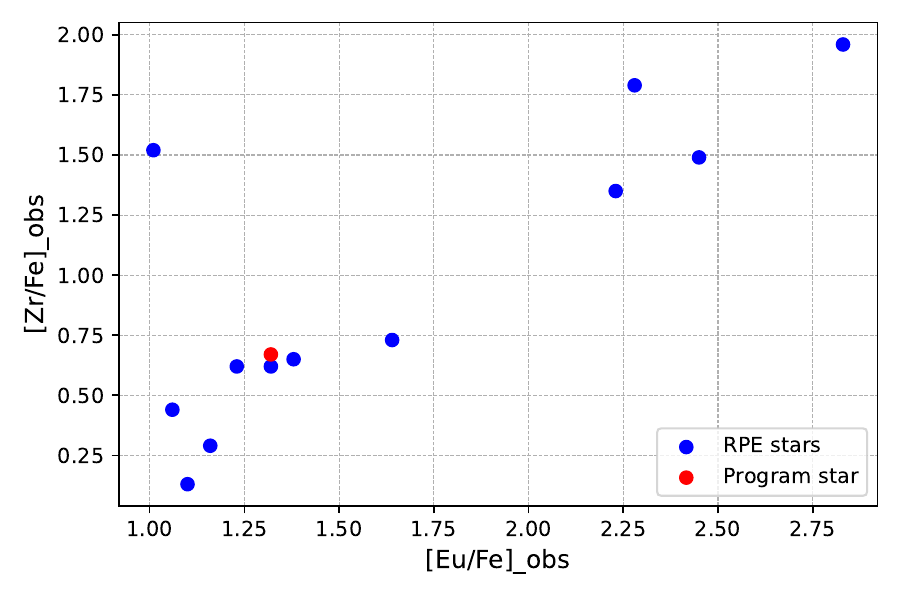}  
        \caption{[Zr/Fe] vs [Eu/Fe]}
        \label{fig:subfig3f}
    \end{subfigure}

    \caption{Comparison of neutron-capture abundance ratios in RPE stars as a function of [Fe/H] and [Eu/Fe]: The blue-filled circles represent RPE stars, and their neutron-capture abundance ratios are plotted against [Fe/H] and [Eu/Fe]. The red-filled circle represents the respective abundance ratio of our program star. These abundance ratios are taken from (\citealt{Roederer+etal+2018,
    Roederer+etal+2024, Ryan+etal+1996, Sneden+etal+2003, Lai+etal+2008, Hayek+etal+2009, Barklem+etal+2005, Aoki+etal+2010, Allen+etal+2012, Li+etal+2015, Xing+etal+2019, Cain+etal+2020}).}
    \label{fig:labe3}
\end{figure}

We predict the abundance ratios and component ratios for this program star to explore the astrophysical origins of the elements. Figure \ref{fig:enter-labe2} compares the observed and calculated abundance ratios, and component ratios.
From Figure \ref{fig:enter-labe2}, we find that the light elements Mg, Si, Ca, and Ti are produced by the primary process, while the Na, Al, Sc, V, Cr, Mn, and Ni are mainly produced by the secondary process of the massive stars. The element Fe is produced by an equal contribution of SNe Ia and the primary process, while Co comes from the mixture of the primary and the secondary processes. The elements Cu and Zn are produced by SNe Ia and the primary process, respectively. In contrast, the heavy $r$-process elements align with the main $r$-process pattern. Because, the main $r$-process is, of course, different from the primary and secondary processes. In general, r-II stars exhibit low [Sr/Ba] ratios. However, some r-I and r-II stars show higher [Sr/Ba] ratios, similar to limited-r stars, even within the same metallicity range \citep{Saraf+etal+2023}. This indicates an enhanced production of Sr in certain $r$-process events. At low metallicity, Ba mainly comes from the $r$-process because the $s$-process, which produces Ba in later stellar evolutionary stages, had not yet contributed significantly. If the same $r$-process produced Sr as Ba, one would expect [Sr/Ba] to exhibit a consistent trend at low metallicities \citep{Travaglio+etal+2004}. However, the significant scatter in [Sr/Ba] with decreasing metallicity suggests that another process may be responsible for Sr production, independent of Ba, as shown in Figure \ref{fig:subfig3a}.

We also examined the behavior of [Sr/Eu] ratios in different highly RPE stars, including our program star, as shown in Figure \ref{fig:subfig3b}. The [Sr/Eu] ratios follow a similar trend to [Sr/Ba], implying that Sr may be produced through processes in addition to the $r$-process, responsible for elements like Europium. Furthermore, correlations between [Sr/Fe], [Y/Fe], and [Zr/Fe] with [Eu/Fe] indicate that some Sr, Y, and Zr are produced in an $r$-process event that synthesizes Eu (see Figures \ref{fig:subfig3d}, \ref{fig:subfig3e}, and \ref{fig:subfig3f}). The significant scatter observed in [Sr/Fe] versus [Eu/Fe] may suggest additional nucleosynthetic sites for Sr production compared to Eu. In contrast, the trends for [Y/Fe] and [Zr/Fe] versus [Eu/Fe] show less scatter, implying that the site producing Sr may not significantly contribute to Y and Zr synthesis. This distinction highlights the complexity of nucleosynthesis and suggests that different processes or events may dominate the production of various neutron-capture elements. The decreasing [Ba/Eu] trend with increasing metallicity shows that the origin of Ba and Eu in our program star is the same (see Figure \ref{fig:subfig3c}). A low [Ba/Eu] ratio is a hallmark to ensure that Ba remains less abundant relative to Eu.

\subsection{Challenges in Fitting Sr, Y, and Zr Abundances: Revisiting $r$-Process Contributions}
The weak and main $r$-processes are believed to occur in violent environments such as CCSNe and NSMs. Two potential pathways for initiating these supernovae have been proposed:

\begin{itemize}
    \item \textbf{Iron Core Collapse:} This happens in stars with progenitor masses $ M\geq 11 M_{\odot}$, where the collapse of the iron core results in the production of both light and Fe-group elements, along with the heavy elements formed via the $r$-process.
    \item \textbf{O-Ne-Mg Core Collapse:} In stars with progenitor masses between 8–10 $M_{\odot}$, the collapse of the oxygen-neon-magnesium core occurs, but it does not produce light or Fe-group elements. 
\end{itemize}

\citet{Li+etal+2013a} constrained the weak $r$-process by analyzing the abundances in weak $r$-process stars HD 122563 and HD 88609, which show minimal contributions from the main $r$-process but relatively high abundances of lighter neutron-capture elements (Sr, Y, Zr). The significant presence of both light and lighter neutron-capture elements in these weak $r$-process stars suggests that they formed from material primarily contaminated by the weak $r$-process. For typical metal-poor stars, the mean value of [Eu/Fe] is approximately 0.3, indicating a slight europium enhancement. However, this ratio shifts in weak $r$-process stars, where pollution from weak $r$-process events results in [Eu/Fe] $\approx -0.5$. This suggests that the processes responsible for producing iron and other light elements in SNe II do not simultaneously yield europium and heavier $r$-process elements in comparable amounts or through similar mechanisms. Additionally, [Sr/Fe] $\approx 0$ in weak $r$-process stars \citep{Honda+etal+2007, Roederer+etal+2012} implies that, although the weak $r$-process does not directly produce iron and light elements, both iron and weak $r$-process elements are ejected together in the same supernova event. Consequently, light, Fe-group, and weak $r$-process elements are considered as a single component, termed the ''primary component''. \citet{Li+etal+2013a} also utilized the abundances of two metal-poor, main $r$-process enhanced stars, CS 22892-052 and CS 31082-001, to calculate the pure main $r$-process components. The resulting pure main $r$-process yields included elements beyond Sr, although Sr itself was also part of the yield. 

\begin{figure*}[!t]
    \centering
    \includegraphics[width=0.49\textwidth]{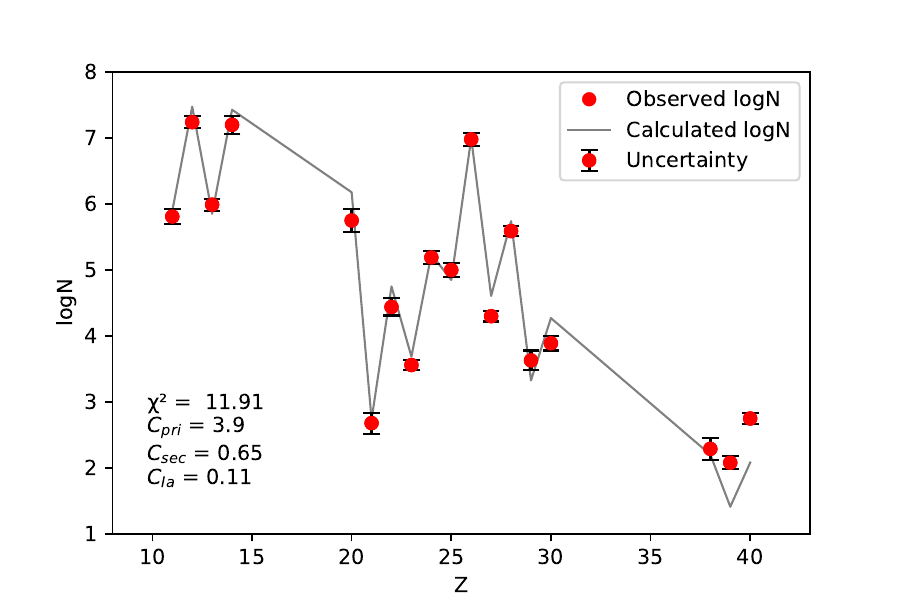}\hfill%
    \includegraphics[width=0.49\textwidth]{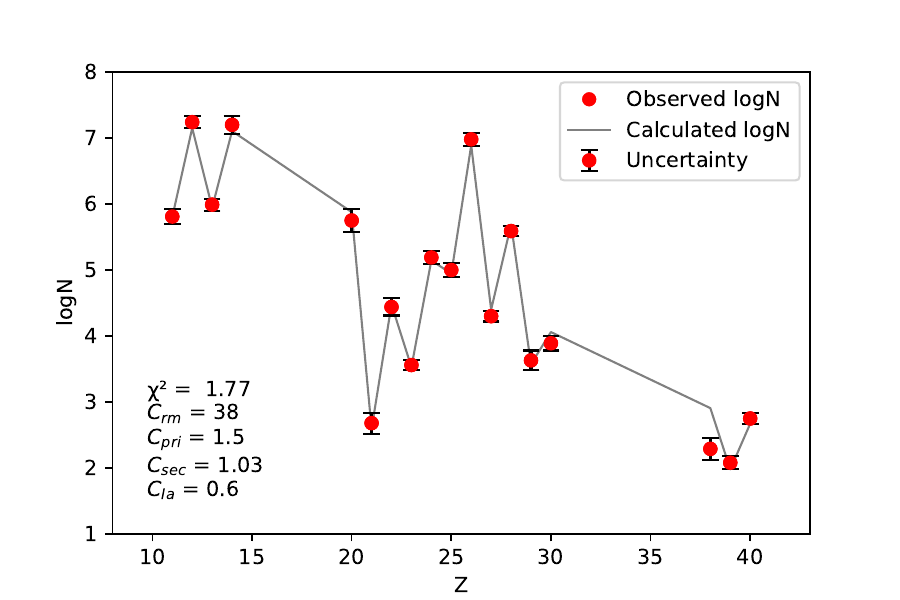}
    \caption{Comparison of the abundance fit with and without the contribution of the main $r$-process in our program star. Left panel: The observed and predicted abundances. The fitting includes only the primary and secondary processes, and SNe Ia contributions (excluding heavy neutron-capture elements). Right panel: similar to the left panel, but the main $r$-process is added during the fitting.}
    \label{fig:enter-labe4}
\end{figure*}
Based on the above observed results, we refit the abundances of our program star, excluding the heavy neutron-capture elements, to examine the contribution of the primary process, specifically for lighter neutron-capture elements (see left panel of Figure \ref{fig:enter-labe4}). Despite modeling this primary process, it can be seen that the observed abundances of Y and Zr remain significantly high. The discrepancy between the observed and calculated abundances of Y and Zr in the program star provides valuable insights into the synthesis processes of neutron-capture elements. This mismatch suggests that the weak $r$-process alone may not fully account for the production of these elements, indicating a potential role for additional processes. While the weak $r$-process generally produces lighter neutron-capture elements such as Sr, Y, and Zr, its limitations become evident in explaining the observed overabundances of Y and Zr in this star.

\begin{figure}[!h]
    \centering
    \begin{subfigure}{0.49\textwidth}
        \centering
        \includegraphics[width=\textwidth]{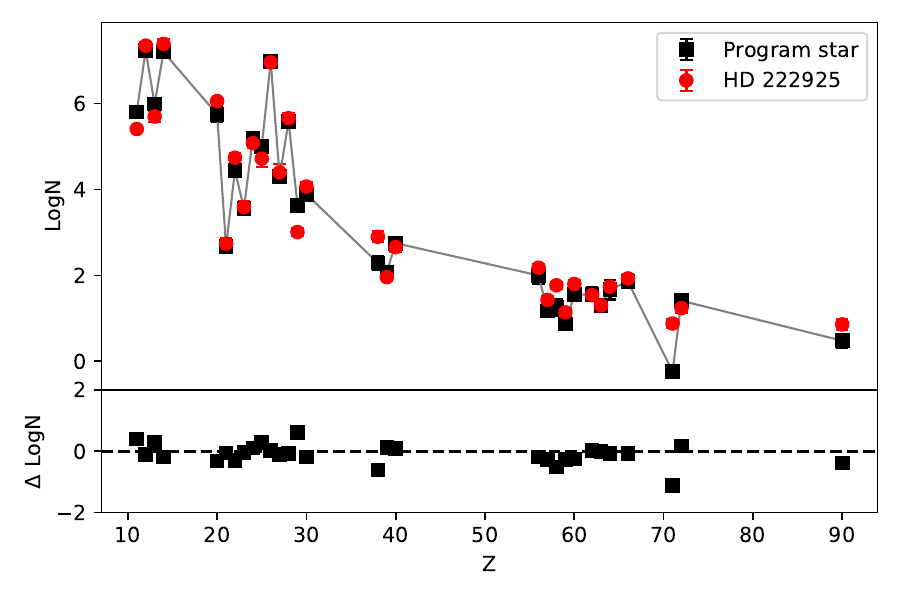}  
        \caption{}
        \label{fig:subfig5a}
    \end{subfigure}
    \begin{subfigure}{0.49\textwidth}
        \centering
        \includegraphics[width=\textwidth]{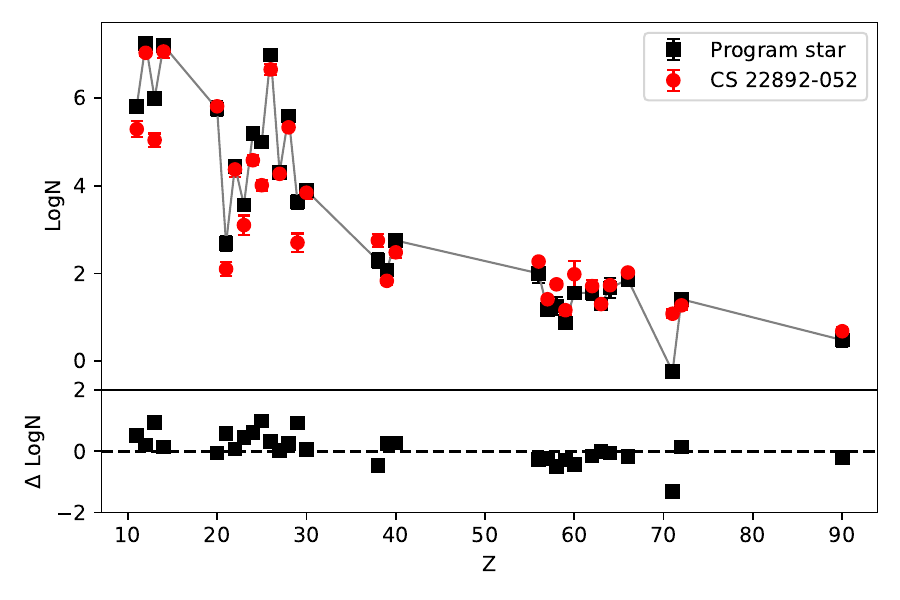}  
        \caption{}
        \label{fig:subfig5b}
    \end{subfigure}
    
    \begin{subfigure}{0.49\textwidth}
        \centering
        \includegraphics[width=\textwidth]{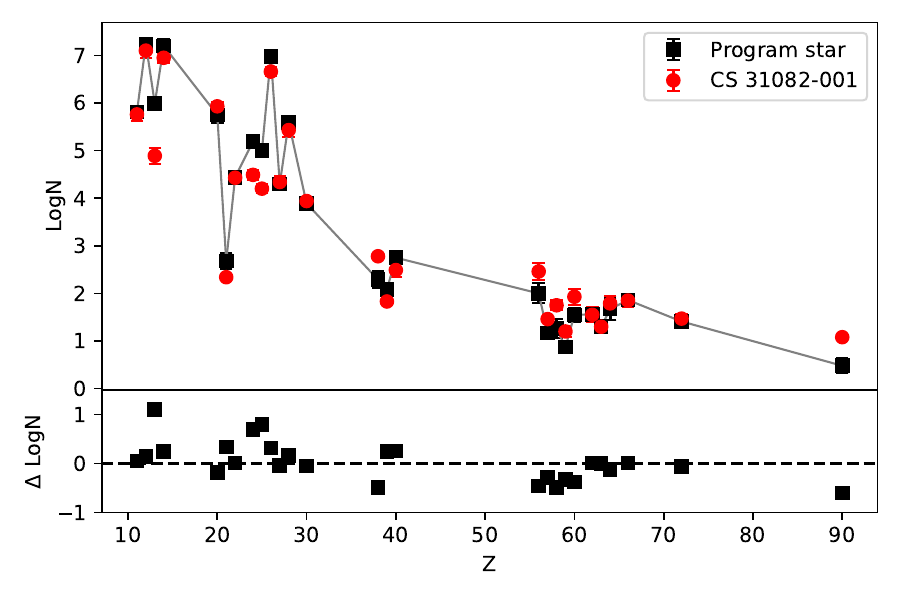}  
        \caption{}
        \label{fig:subfig5c}
    \end{subfigure}
    \begin{subfigure}{0.49\textwidth}
        \centering
        \includegraphics[width=\textwidth]{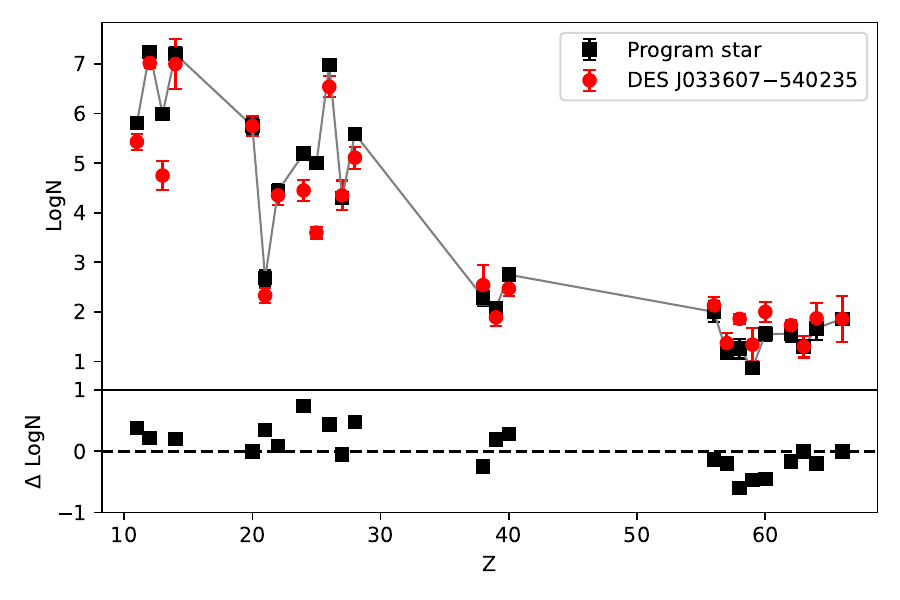}  
        \caption{}
        \label{fig:subfig5d}
    \end{subfigure}
    \caption{Top panels: Comparison of the observed abundances of our program star with other known r-II stars. The observed abundances of the program star are shown as black squares, while those of other r-II stars are represented by red circles. Bottom panels: The relative offsets (\(\Delta \log N = \log N_{\text{program star}} - \log N_{\text{r-II stars}}\)) for the selected elements.}
    \label{Fig:enter-labe5}
\end{figure}
The excess production of Y and Zr compared to theoretical models indicates that the environment of this star was particularly favorable for the synthesis of these elements. The results suggest that mechanisms other than the weak $r$-process may have played a substantial role in their creation. Such overabundances could stem from a stronger $r$-process contribution, a secondary $r$-process, or even exotic nucleosynthesis events not included in current models. This finding highlights the complex nucleosynthetic history of this program star and suggests that its elemental enrichment likely resulted from a combination of processes. We examined the behavior of lighter neutron-capture elements in the program star by incorporating the main $r$-process component (see right panel of Figure \ref{fig:enter-labe4}). The observed abundances of Y and Zr align closely with the predicted values, indicating that the main $r$-process significantly contributes to the synthesis of these elements. However, the discrepancy between the observed and calculated abundances of Sr remains an open question regarding its origin. Furthermore, from Figure \ref{fig:enter-labe1} we observe the difference between observed [Sr/Eu]$_{\text{obs}} = -1.4$ and predicted main $r$-process value of [Sr/Eu]$_{\text{m,r}} = -1.03$ which is -0.37. It may indicate that the main $r$-process model overestimates the Sr yield or that post-nucleosynthesis processes have reduced the Sr abundance in the program star. This slight overestimation could hint at the need for refining $r$-process models or investigating additional astrophysical mechanisms affecting Sr retention or dilution. In contrast, the observed ratios [Y/Eu]$_{\text{obs}} = -0.91$ and [Zr/Eu]$_{\text{obs}} = -0.65$ are higher than their respective calculated values from the main $r$-process, which are [Y/Eu]$_{\text{m,r}} = -1.27$ and [Zr/Eu]$_{\text{m,r}} = -0.97$. This suggests that while the main $r$-process contributes to the synthesis of lighter neutron-capture elements, the adopted main $r$-process model does not adequately fit the observed data. Therefore, another main $r$-process pattern may exist.

\section{Analysis on the Formation and Evolutionary history of the Program star}
\label{sec:3}
We compare the observed abundances of our program star with those of other known r-II stars. HD 222925, CS 22892-052, and CS 31082-001 belong to the MW halo, while DES J033607-540235 is a member of the Reticulum II dwarf galaxy (\citealt{Roederer+etal+2018, Sneden+etal+2008, Ji+etal+2016}). The observed abundances of the r-II stars are normalized to the Eu abundance of our program star. We select elements from the r-II stars similar to our program star for better comparison. The results are shown in Figure \ref{Fig:enter-labe5}. It can be seen that the abundance pattern for the heavy neutron-capture elements (\(Z\geq56 \)) exhibits less deviation compared to the light elements, suggesting that the main $r$-process is more consistent for these elements. \citet{Roederer+etal+2018} explained that the main $r$-process material of HD 222925 is influenced by a single, high-yield  $r$-process source, potentially linked to an NSM. We find that the abundance of Lu is exceptionally low in our program star compared to HD 222925 and CS 22892-052, as is evident in Figure \ref{fig:enter-labe1} also. This difference may suggest a more complex enrichment history or variation in the yields of the contributing $r$-process site(s).
Theoretical models and observational evidence suggest that NSMs play a major role in enriching $r$-process elements in the Galactic halo, though other sources such as MRSNe or collapsars may contribute at lower metallicities (\citealt{ Cote+etal+2019, Cowan+etal+2021}). The unique abundance pattern identified in our program star, especially the underabundance of mid-range heavy elements like Ce, Pr, and Nd compared to the scaled solar $r$-process pattern, offers essential insights into its nucleosynthetic origins. While promoting the production of heavier nuclei like Eu, physical circumstances such as modest neutron richness or enhanced neutrino flow restricted the synthesis of these elements. The most likely cause of these deviations is a non-uniform $r$-process event, which might be linked to the MRSNe or an NSM (\citealt{vandeVoort+etal+2020}). Due to the strange $r$-process pattern observed in our program star, the exact site of the $r$-process is uncertain.

Additionally, the production of mid-range $r$-process elements would have been hindered by the accretion disk winds in an NSM, which normally eject material with greater electron fractions (\citealt{Just+etal+2015}). Our program star's location within the thin disk indicates that its chemical development was driven by specialized $r$-process contributions, with little interaction with typical $r$-process sources. As a result, the abundance pattern of our program star exemplifies the variability and unpredictability of elemental yields created by various $r$-process conditions.
\section{Conclusion}
\label{sec:4}

We have conducted a comprehensive analysis of the astrophysical sources of the neutron-capture elements in LAMOST J020623.21+494127.9. Using the abundance decomposition method, we fit the abundances of 25 elements in this star, in which 10 elements are heavy neutron-capture elements. We derived the five component coefficients which are \( C_{r,m} = 18.05 \), \( C_{\text{pri}} = 1.36 \), \( C_{s,m} = 0 \), \( C_{\text{sec}} = 1.18 \), and \( C_{\text{Ia}} = 0.56 \). Our results are given as follows:

The decomposition results of our program star reflect the astrophysical processes involved in its formation. Since it is an r-II star, our analysis shows that heavy neutron-capture elements are predominantly produced by the main $r$-process. Additionally, we find that the $\alpha$-elements (Mg, Si, and Ca) mainly originate from the primary process, while light elements such as Na, Al, Sc, Cr, Mn, and Ni are primarily synthesized through the secondary process. This indicates that massive stars play a significant role in the production of light elements in the program star. The element Fe is produced by an equal contribution of SNe Ia and the primary process, while Cu and Zn are produced by SNe Ia and the primary process, respectively.

The discrepancy between the observed and calculated abundances of Y and Zr in the program star suggests that the adopted weak $r$-process pattern alone cannot fully account for the production of these elements, indicating the involvement of additional nucleosynthetic processes. By examining the behavior of lighter neutron-capture elements in the fitting process, excluding heavy neutron-capture elements, and incorporating the main $r$-process component, we find that the observed abundances of Y and Zr align well with the predicted values, highlighting the significant contribution of the main $r$-process to their synthesis. Figure~\ref{fig:enter-labe1} shows that the difference between observed ratio $[\text{Sr}/\text{Eu}]_{\text{obs}} = -1.4$ and the predicted main $r$-process value $[\text{Sr}/\text{Eu}]_{\text{m,r}} = -1.03$ is -0.37. The slight overestimation of Sr by the main $r$-process could hint at the need for refining these models or investigating additional astrophysical mechanisms affecting Sr retention or dilution. However, the observed ratios $[\text{Y}/\text{Eu}]_{\text{obs}}$ and $[\text{Zr}/\text{Eu}]_{\text{obs}}$ exceed the calculated values for the main $r$-process. This suggests that while the main $r$-process contributes to the synthesis of lighter neutron-capture elements, the adopted main $r$-process model does not adequately explain the observed data, pointing to the potential existence of an alternative main $r$-process pattern.

The unique abundance pattern observed in the program star, particularly the underabundance of mid-range heavy elements like Ce, Pr, and Nd compared to the scaled solar $r$-process pattern, suggests that physical conditions such as moderate neutron richness or increased neutrino flux may have limited the synthesis of these elements while facilitating the formation of heavier nuclei like Eu. These deviations are likely the result of a non-uniform $r$-process event, potentially associated with an NSM or MRSNe.

\begin{acknowledgements}
This study is supported by the National Key Basic R$\&$D Program of China No. 2024YFA1611903, the National Natural Science Foundation of China under grant No. 12173013, the project of Hebei provincial department of science and technology under the grant number 226Z7604G, and the Hebei NSF (No. A2021205006). C.W. acknowledges the China Manned Space Project for funding support of this study.  

\end{acknowledgements}
  
\bibliographystyle{raa}
\bibliography{bibtex}++

\end{document}